# Symmetry-breaking bifurcations of pure-quartic solitons in dual-core couplers

PENGFEI LI,[1,2,*] LIANGLIANG DONG,[1,2] DUMITRU MIHALACHE[3], AND BORIS A. MALOMED[4,5]

[1]*Department of Physics, Taiyuan Normal University, Jinzhong, 030619, China*
[2]*Institute of Computational and Applied Physics, Taiyuan Normal University, Jinzhong, 030619, China*
[3]*Horia Hulubei National Institute of Physics and Nuclear Engineering, Magurele, Bucharest RO-077125, Romania*
[4]*Department of Physical Electronics, School of Electrical Engineering, Faculty of Engineering, and Center for Light-Matter Interaction, Tel Aviv University, Tel Aviv 69978, Israel*
[5]*Instituto de Alta Investigación, Universidad de Tarapacá, Casilla 7D, Arica, Chile*
*Corresponding author: lpf281888@gmail.com*

**We investigate spontaneous symmetry- and antisymmetry-breaking bifurcations of solitons in a nonlinear dual-core waveguide with the pure-quartic dispersion and Kerr nonlinearity. Symmetric, antisymmetric, and asymmetric pure-quartic solitons (PQSs) are found, and their stability domains are identified. The bifurcations for both the symmetric and antisymmetric PQSs are of the supercritical type (alias phase transitions of the second kind). Direct simulations of the perturbed evolution of PQSs corroborate their stability boundaries predicted by the analysis of small perturbations.** © 2024 Optical Society of America

http://dx.doi.org/xx.xxxx/OL.xx.xxxxxx

In nonlinear optical waveguides with the intrinsic double-well structure, a generic effect that occurs with the increase of the nonlinearity strength is the spontaneous symmetry breaking (SSB). This effect was first considered, in a mathematical form, in the framework of the nonlinear Schrödinger equation (NLSE) with the self-focusing nonlinearity [1]. Then, it was predicted, as physical effect, in the models of an optically-induced photonic lattice [2]. A ramification of the theme is SSB in dual-core systems, such as nonlinear optical couplers. In such systems, the SSB phase transitions were studied in detail theoretically [3-7], and demonstrated experimentally in a twin-core fiber [8].

In the framework of the NLSE, the studies of SSB of optical solitons have been expanded in other directions, one of which deals with *PT*-symmetric NLSEs that model nonlinear optical waveguides with balanced distributions of gain and loss [9]. In that setting, SSB of optical solitons occurs in a class of complex *PT*-symmetric double-well potential [10-16]. On the other hand, optical solitons have been intensively studied in the framework of fractional NLSEs, which include terms representing fractional-order diffraction or dispersion [17-19]. In the framework of the fractional NLSE, SSB of optical solitons has been explored under the action of real [20-22] and *PT*-symmetric [23-26] potentials. The latter setting has initiated systematic studies of a novel type of asymmetric soliton modes, *viz.*, ghost states with complex propagation constants [23,24].

Recently, much interest was drawn to a new variety of optical solitary waves in the form of pure-quartic solitons (PQSs), which exists in nonlinear media with the fourth-order dispersion (FOD) [27]. They have been experimentally created by means of the precise dispersion engineering [28]. PQSs feature oscillations in their exponentially decaying tails, which makes them drastically different from conventional solitons in optical media with the quadratic dispersion [29]. In the framework of the NLSE with the FOD, various species of solitons have been predicted, including quartic dissipative solitons [30,31], Raman PQSs [32,33], and dark solitons [34-36]. PQSs have also been experimentally demonstrated in reconfigurable mode-locked fiber lasers [37,38]. However, the SSB phenomenology and asymmetric PQSs in coupled NLSEs with dominant FOD terms, which is the subject of the present work, have not been considered previously.

As the basic model, we use a system of coupled NLSEs for amplitudes $\Psi_1$ and $\Psi_2$ of optical waves, under the combined effects of FOD, self-focusing Kerr nonlinearity, and linear coupling:

$$i\frac{\partial \Psi_{1,2}}{\partial z} - \frac{1}{24}\frac{\partial^4 \Psi_{1,2}}{\partial \tau^4} + |\Psi_{1,2}|^2 \Psi_{1,2} + \Psi_{2,1} = 0, \quad (1)$$

where $z$ is the propagation distance and $\tau$ is the retarded time in the reference frame moving with the carrier group velocity. Stationary solutions to Eq. (1) with a real propagation constant $\beta$ are sought for as

$$\Psi_{1,2}(\tau, z) = \psi_{1,2}(\tau) e^{i\beta z}, \quad (2)$$

where $\psi_1$ and $\psi_2$ satisfy the coupled equations,

$$-\frac{1}{24}\frac{d^4 \psi_{1,2}}{d\tau^4} + |\psi_{1,2}|^2 \psi_{1,2} - \beta \psi_{1,2} + \psi_{2,1} = 0. \quad (3)$$

Soliton solutions of Eq. (3) are characterized by the total power, which is a dynamical invariant of Eq. (1), defined as

$$P(\beta) = \int_{-\infty}^{+\infty} \left(|\psi_1|^2 + |\psi_2|^2\right) d\tau. \quad (4)$$

To produce numerical soliton solutions of Eq. (3), we employed the Newton-conjugate-gradient (NCG) method [39], applying it to the coupled NLSEs. The stability of the solitons was addressed by performing the linearization procedure for eigenmodes $u_{1,2}(\tau)$ and $v_{1,2}(\tau)$ of small perturbations, with eigenvalue $\delta \equiv \delta_R + i\delta_I$, $\delta_R$ being the linear-instability growth rate, $\delta_R = 0$ indicating that the underlying stationary solutions $\psi_1$ and $\psi_2$ are linearly stable. Thus, the perturbed solution is taken as

$$\Psi_1(\tau,z) = \left[\psi_1(\tau) + u_1(\tau)e^{\delta z} + u_2^*(\tau)e^{\delta^* z}\right]e^{i\beta z}, \quad (5)$$

$$\Psi_2(\tau,z) = \left[\psi_2(\tau) + v_1(\tau)e^{\delta z} + v_2^*(\tau)e^{\delta^* z}\right]e^{i\beta z}. \quad (6)$$

Substituting it in Eq. (1) and performing the linearization, we arrive at the eigenvalue problem

$$i\left[\left(-\frac{1}{24}\frac{d^4}{d\tau^4} - \beta + 2|\psi_1|^2\right)u_1 + \psi_1^2 u_2 + v_1\right] = \delta u_1, \quad (7)$$

$$i\left[\left(+\frac{1}{24}\frac{d^4}{d\tau^4} + \beta - 2|\psi_1|^2\right)u_2 - \psi_1^{*2} u_1 - v_2\right] = \delta u_2, \quad (8)$$

$$i\left[\left(-\frac{1}{24}\frac{d^4}{d\tau^4} - \beta + 2|\psi_2|^2\right)v_1 + \psi_2^2 v_2 + u_1\right] = \delta v_1, \quad (9)$$

$$i\left[\left(+\frac{1}{24}\frac{d^4}{d\tau^4} + \beta - 2|\psi_2|^2\right)v_2 - \psi_2^{*2} v_1 - u_2\right] = \delta v_2. \quad (10)$$

We first obtained PQS families, produced by the application of the NCG method to Eq. (3). Then, the spectrum of stability eigenvalues $\delta$ was produced by the numerical solution of Eqs. (7)-(10) by means of the Fourier collocation method [39].

The numerical solution reveals four types of PQS families, viz., symmetric and antisymmetric ones, whose propagation constants are, respectively, $\beta = 1$ and $\beta = -1$ in the linear limit, and two species of asymmetric solutions produced by SSB bifurcations.

The family of the asymmetric PQSs of the first type is represented, in Fig. 1(a), by the dependence of their total power on the propagation constant. The family of the asymmetric PQSs of the first type is created by the SSB bifurcation of the basic symmetric-PQS family at the critical point $P_{cr1} \approx 3.35$, $\beta_{cr1} \approx 1.82$.

At $\beta > \beta_{cr1}$, the symmetric solitons are unstable, as their maximum linear-instability growth rate is positive ($\max(\text{Re}(\delta)) > 0$), while the family of the asymmetric solitons of the first type is stable, with $\text{Re}(\delta) = 0$, in their entire existence domain. Generic examples of the symmetric and asymmetric PQSs are displayed, respectively, in Figs. 1(b) and 1(c) on the linear scale, and in Figs. 1(d) and 1(e) on the logarithmic one, for $\beta = 3$. In contrast to the smooth hyperbolic-secant-like shapes of conventional symmetric and asymmetric solitons in the system with the quadratic dispersion [6,7], PQSs feature oscillations in exponentially decaying tails, which is a manifestation of FOD [29].

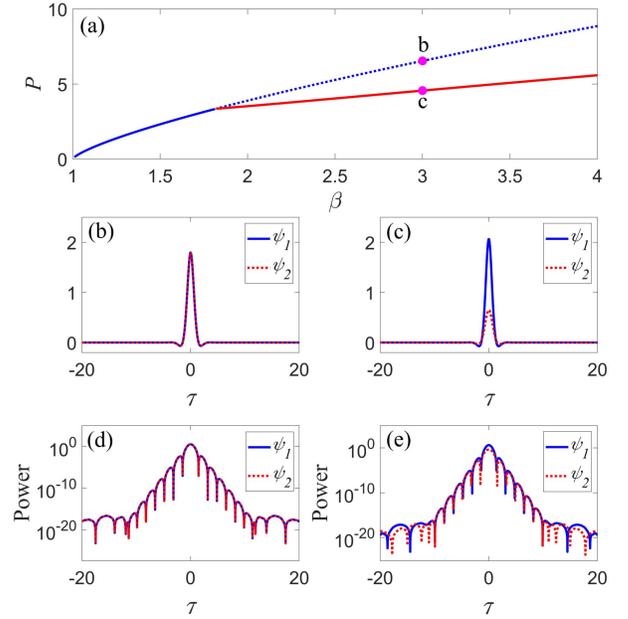

Fig. 1. (a) The blue solid (dashed) lines plot dependences of the PQS power $P$ on the propagation constant $\beta$ for stable (unstable) symmetric solitons. The red solid line is the same dependence for stable asymmetric solitons of the first type. Panels (b) and (c) display, severally, examples of the unstable symmetric PQS and stable asymmetric one of the first type, produced as numerical solutions of Eq. (3) with $\beta = 3$, marked by the corresponding magenta dots in panel (a). Panels (d) and (e) show the same as (b) and (c), but on the logarithmic scale, revealing oscillatory tails of the solitons.

Asymmetric PQSs of the second type are introduced in Fig. 2(a). As shown in the figure, they are produced by the bifurcation of the spontaneous *antisymmetry breaking*. It is shown too that the antisymmetric PQSs are stable in a very narrow region, $\beta < -0.88$, while the antisymmetry-breaking bifurcation occurs at point $P_{cr2} \approx 6.61$, $\beta_{cr2} \approx 1.03$. The entire family of asymmetric PQSs of the second type is unstable as it is created by the bifurcation from the unstable antisymmetric states. Examples of the unstable antisymmetric PQS and one with the spontaneously broken antisymmetry are presented, respectively in Fig. 2(b) and 2(c) on the linear scale, and in Figs. 2(d) and 2(e) on the logarithmic one.

The symmetry- and antisymmetry- breaking phase transitions of the PQS families are further characterized in Fig. 3 by means of plots of the asymmetry parameter $\Theta$ vs. the propagation constant. The asymmetry is defined as

$$\Theta = \frac{P_1 - P_2}{P_1 + P_2}, \quad (11)$$

where the integral powers are

$$P_{1,2}(\beta) = \int_{-\infty}^{+\infty} |\psi_{1,2}(\tau)|^2 d\tau. \quad (12)$$

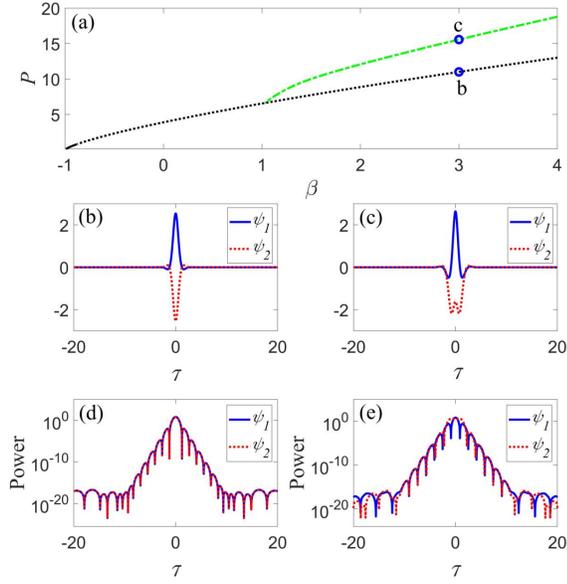

Fig. 2. (a) The same as in Fig. 1(a), but for the antisymmetric PQSs and asymmetric PQSs of the second type, which are generated from the (unstable) antisymmetric ones by the antisymmetry-breaking bifurcation. Black solid (dashed) lines indicate families of stable (unstable) antisymmetric solitons, while the green dashed line designates the completely unstable branch of the solitons with the broken antisymmetry. Examples of the unstable antisymmetric and antisymmetry-broken PQSs are displayed in panels (b) and (c), for the propagation constant $\beta = 3$, as marked, respectively, by the corresponding blue circles in panel (a). Panels (d) and (e) show the same as (b) and (c), but on the logarithmic scale, revealing oscillatory tails of the solitons.

The SSB bifurcations in Figs. 3(a) and 3(b) are of the supercritical type [40], i.e., they are symmetry/antisymmetry-breaking phase transitions of the second kind. This is in contrast with the conventional system with the second-order dispersion, where the symmetry- and antisymmetry-breaking soliton bifurcations are of the subcritical and supercritical types, respectively [6,7].

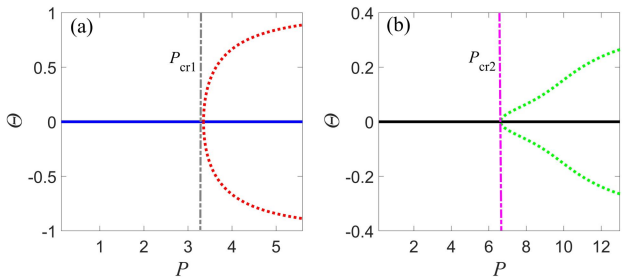

Fig. 3. Asymmetry parameter (11) vs. total power (3). (a) The SSB bifurcations relating the symmetric solitons (the blue solid line) and stable asymmetric ones of the first type (the red dotted line). (b) The antisymmetry-breaking bifurcation relating the antisymmetric solitons (the black solid line) and unstable asymmetric ones of the second type (the green dotted line). The vertical dashed-dotted lines designate the corresponding bifurcation points.

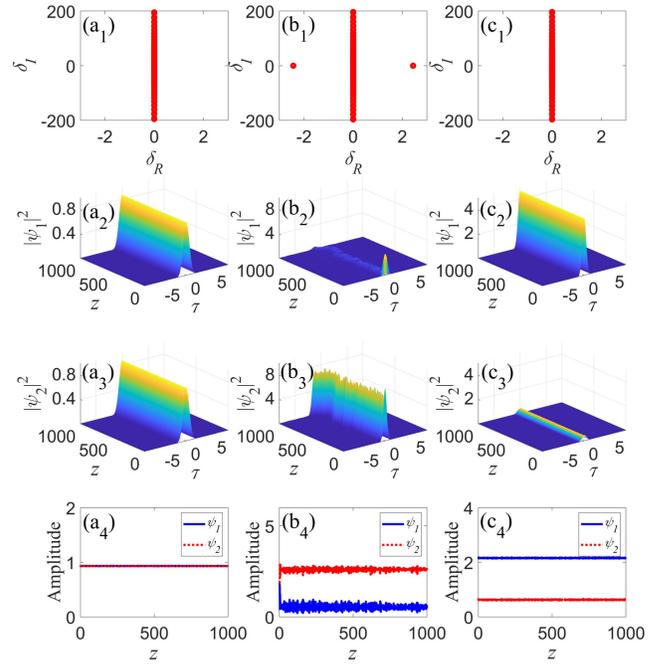

Fig. 4. The linear-stability spectra and evolution of the symmetric and the first type of asymmetric QPSs. The left-hand column shows the stable symmetric soliton with $\beta = 1.5$ and random perturbations at a 5% amplitude level: the linear-stability spectrum ($a_1$), intensities $|\psi_1|^2$ ($a_2$) and $|\psi_2|^2$ ($a_3$), and the soliton's amplitudes (the maximum values of $|\psi_{1,2}|$) vs. the propagation distance ($a_4$). Middle ($b_1, b_2, b_3, b_4$) and right-hand ($c_1, c_2, c_3, c_4$) columns show the same for unstable symmetric and stable asymmetric solitons, taken at $\beta = 3$.

Next, we examine the stability and evolution of the two-component PQSs, by means of the numerical solution of the eigenvalue problem (7)-(10), parallel to direct simulations of the perturbed evolution. The results for symmetric and symmetry-broken solitons, as well as for the antisymmetric and antisymmetry-broken ones, are presented in Figs. 4 and 5, respectively. In all cases, unstable eigenvalues are purely real, and evolution of the instability leads to chaotic oscillations and possible splitting of the solitons.

The stability of the symmetric PQS below the SSB bifurcation point (at $\beta < \beta_{cr1} \approx 1.82$) is illustrated by Figs. 4($a_1$)-4($a_4$). The results indicate that the symmetric soliton (taken at $\beta = 1.5$) remain stable at least up to $z = 1000$ under the action of random initial perturbations with a relative amplitude of 5%. Beyond the bifurcation point, the symmetric soliton is unstable, while the asymmetric one is stable. The corresponding linear-stability spectra are shown in Figs. 4($b_1$) and 4($c_1$). Direct simulations shown in Figs. 4($b_2$)-4($b_4$), demonstrate that the unstable symmetric soliton quickly evolves to an asymmetric shape. On the other hand, Figs. 4($c_2$)-4($c_4$) corroborate the robust propagation of the stable asymmetric soliton.

Outcomes of the evolution of the antisymmetric antisymmetry-broken solitons are presented in Figs. 5. Linearly stable and unstable antisymmetric solitons are plotted in Figs. 5($a_1$) and 5($b_1$), respectively. The simulations, shown in Figs. 5($a_2$)-5($a_4$), and 5($b_2$)-5($b_4$) corroborate, severally, their stability and instability (the onset of the instability, leading to the emergence of randomly evolving modes, is readily observed even in the absence of random perturbations). Figures 5($c_1$)-5($c_4$) demonstrate that the unstable asymmetric soliton of the second type (the one with the broken antisymmetry) survives as a quasi-stable mode at the early stage of the evolution ( $z < 10$ ), which is followed by the onset of conspicuous instability.

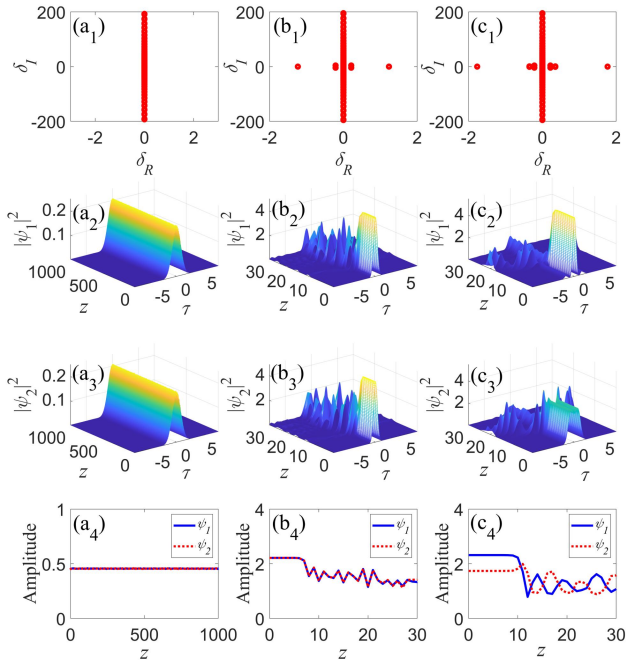

Fig. 5. The linear-stability spectra and evolution for the antisymmetric solitons and ones with broken antisymmetry (cf. Fig. 4 for the symmetric and asymmetric PQSs). The left-hand column ($a_1$-$a_4$) presents a stable antisymmetric PQS at $\beta = -0.88$. The middle ($b_1$-$b_4$) and right-hand ($c_1$-$c_4$) columns show the unstable antisymmetric soliton and one with broken antisymmetry at $\beta = 2$.

In conclusion, we have reported the scenario of SSB (spontaneous symmetry breaking) for the PQSs (pure quartic solitons) in the dual-core optical waveguide with FOD (fourth-order dispersion). Two types of asymmetric QPSs are produced by the SSB bifurcation from the symmetric and antisymmetric solitons. In contrast with the conventional system with the second-order dispersion, the SSB bifurcations are identified here as supercritical ones. Stability domains have been identified for the symmetric, antisymmetric, and asymmetric PQSs through the calculation of the respective small-perturbation spectra, and verified by direct simulations of the perturbed evolution.

As an extension of the work, it may be relevant to expand the analysis for *PT*-symmetric nonlinear coupler and two-dimensional version of the system with the pure-quartic dispersion.

**Funding.** The work of P.F.L. was supported by the National Natural Science Foundation of China (NNSFC) (11805141) and Applied Basic Research Program of Shanxi Province (202203021222250, 202303021211185). The work of B.A.M. is supported by the Israel Science Foundation (grant No. 1695/22).

**Disclosures.** The authors declare no conflicts of interest.

**Data Availability.** Data underlying the results presented in this paper are not publicly available at this time but may be obtained from the authors upon reasonable request.

## REFERENCES

1. E. B. Davies, Commun. Math. Phys. **64**, 191 (1979).
2. J. C. Eilbeck, P. S. Lomdahl, and A. C. Scott, Physica D **16**, 318 (1985).
3. E. M. Wright, G. I. Stegeman, and S. Wabnitz, Phys. Rev. A **40**, 4455 (1989).
4. C. Paré and M. Fłorjańczyk, Phys. Rev. A **41**, 6287 (1990).
5. A. W. Snyder, D. J. Mitchell, L. Poladian, D. R. Rowland, and Y. Chen, J. Opt. Soc. Am. B **8**, 2102 (1991).
6. N. Akhmediev and A. Ankiewicz, Phys. Rev. Lett. **70**, 2395 (1993).
7. B. A. Malomed, I. Skinner, P. L. Chu, and G. D. Peng, Phys. Rev. E **53**, 4084 (1996).
8. V. H. Nguyen, L. X. T. Tai, I. Bugar, M. Longobucco, R. Buczyński, B. A. Malomed, and M. Trippenbach, Opt. Lett. **45**, 5221 (2020).
9. V. V. Konotop, J. Yang, and D. A. Zezyulin, Rev. Mod. Phys. **88**, 035002 (2016).
10. J. Yang, Opt. Lett. **39**, 5547 (2014).
11. P. F. Li, D. Mihalache, and L. Li, Rom. J. Phys. **61**, 1028 (2016).
12. P. F. Li and D. Mihalache, Proc. Rom. Acad. A **19**, 61 (2018).
13. P. F. Li, C. Q. Dai, R. J. Li, and Y. Q. Gao, Opt. Express **26**, 6949 (2018).
14. J. K. Yang, Opt. Lett. **44**, 2641 (2019).
15. J. K. Yang, Phys. Rev. E **91**, 023201 (2015).
16. L. W. Dong, C. M. Huang, and W. Qi, Nonlinear Dyn. **98**, 1701 (2019).
17. B. A. Malomed, Photonics **8**, 353 (2021).
18. P. G. Kevrekidis and J. Cuevas-Maraver, *Fractional Dispersive Models and Applications* (Springer, 2024).
19. D. Mihalache, Rom. Rep. Phys. **76**, 402 (2024).
20. P. F. Li, B. A. Malomed, and D. Mihalache, Chaos, Solitons Fractals **132**, 109602 (2020).
21. P. F. Li and C. Q. Dai, Ann. Phys. (Berlin) **532**, 2000048 (2020).
22. P. F. Li, R. J. Li, and C. Q. Dai, Opt. Express **29**, 3193 (2021).
23. P. F. Li, B. A. Malomed, and D. Mihalache, Opt. Lett. **46**, 3267 (2021).
24. M. Zhong, L. Wang, P. F. Li, and Z. Y. Yan, Chaos **33**, 013106 (2023).
25. M. Zhong and Z. Y. Yan, Commun. Phys. **6**, 92 (2023).
26. X. Q. He, Y. B. Zhai, Q. Cai, R. J. Li, and P. F. Li, Chaos, Solitons Fractals **186**, 115258 (2024).
27. C. M. de Sterke, A. F. Runge, D. D. Hudson, and A. Blanco-Redondo, APL Photonics **6**, 091101 (2021).
28. A. Blanco-Redondo, C. M. de Sterke, J. E. Sipe, T. F. Krauss, B. J. Eggleton, and C. Husko, Nature Commun. **7**, 10427 (2016).
29. K. K. K. Tam, T. J. Alexander, A. Blanco-Redondo, and C. M. de Sterke, Opt. Lett. **44**, 3306 (2019).
30. H. Taheri and A. B. Matsko, Opt. Lett. **44**, 3086 (2019).
31. Z. C. Qian, M. Liu, A. P. Luo, Z. C. Luo, and W. C. Xu, Opt. Express **30**, 22066 (2022).
32. K. Liu, S. Yao, and C. Yang, Opt. Lett. **46**, 993 (2021).
33. Z. T. Wang, C. Y. Luo, Y. W. Wang, X. H. Ling, and L. F. Zhang, Opt. Lett. **47**, 3800 (2022).
34. T. J. Alexander, G. A. Tsolias, A. Demirkaya, R. J. Decker, C. M. de Sterke, and P. G. Kevrekidis, Opt. Lett. **47**, 1174 (2022).
35. P. Parra-Rivas, S. Hetzel, Y. V. Kartashov, P. F. de Córdoba, J. A. Conejero, A. Aceves, and C. Milián, Opt. Lett. **47**, 2438 (2022).
36. P. Gao, L. Lv, and Xin Li, Opt. Express **32**, 19517 (2024).
37. A. F. J. Runge, D. D. Hudson, K. K. K. Tam, C. Martijn de Sterke, and A. Blanco-Redondo, Nat. Photonics **14**, 492 (2020).
38. J. P. Lourdesamy, A. F. J. Runge, T. J. Alexander, D. D. Hudson, A. Blanco-Redondo, and C. Martijn de Sterke, Nat. Phys. **18**, 59 (2022).
39. J. K. Yang, *Nonlinear Waves in Integrable and Nonintegrable Systems* (SIAM, 2010).
40. G. Iooss and D. D. Joseph, *Elementary Stability Bifurcation Theory* (Springer: New York, 1980).